\documentstyle[12pt]{article}
\newcommand{\lm}[1]{\mbox{$\lambda_{#1}$}}
\newcommand{\m}[1]{\mbox{$m_{#1}^{2}$}}
\newcommand{\p}[1]{\mbox{$\phi_{#1}$}}
\renewcommand{\H}[1]{\mbox{$H_{#1}$}}
\newcommand{\Hd}[1]{\mbox{$H_{#1}^{\dag}$}}
\newcommand{\s}{\mbox{supersymmetry }}

\begin{document}
\setcounter{page}{0}
\thispagestyle{empty}
\begin{flushright}
GUTPA/95/11/1
\end{flushright}
\vskip .2in
\begin{center}
{\LARGE\bf Electroweak Baryogenesis  \\
in the Next to Minimal \\
\vskip .1in
Supersymmetric Model\footnote{Research partially
 funded by the UK Particle Physics and Astronomy Research Council}}
\vskip 1.2cm
{\large\bf A. T. Davies, C. D. Froggatt and R. G. Moorhouse\\}
\vskip .4cm
{ \em Department of Physics and Astronomy,
\\University of Glasgow
\\Glasgow G12 8QQ, U.K.}
\end{center}
\section*{ }
\begin{center}
{\large\bf Abstract}
\end{center}
In the electroweak phase transition there arises the problem of baryon
number washout by sphaleron transitions, which can be avoided if the phase
transition is strongly enough first order. The minimal supersymmetric standard
model has just two Higgs doublets \H{1} and \H{2}, while the next to minimal
model, NMSSM, has an additional singlet, N, this latter giving rise to the
 helpful feature that the Higgs potential contains a tree level trilinear
 field term. We use the tunneling criterion for the existence of
 a first order electroweak phase change. A quantitative statistical analysis
 indicates that with parameters of the NMSSM satisfying the experimental
 constraints a strong first order phase change occurs in about 50\% of
 cases.

\pagebreak

\section{Introduction} \label{intro}
 It is well known that there is difficulty in
sustaining the hypothesis \cite{Kuzm} of
baryogenesis at the electroweak phase
transition in the minimal standard model.
In this standard model one Higgs
doublet case the source of CP violation is
the CKM quark mixing matrix which is too small to
explain the observed baryon to entropy ratio.
Also, requirements on the phase
transition (in a perturbative treatment)
seem to lead to a Higgs mass smaller
than the experimental lower limit \cite{Dine,Boch}.
To overcome these difficulties,
attention has been given to extensions of
the minimal Standard Model, involving
the addition of extra scalars
\cite{Boch,Turk,Andn,Din2,Cohn,Davl}. Prominent among
these is the extension to two Higgs doublets;
other rather natural such
extensions, necessitating two Higgs doublets,
are supersymmetric theories.
In the minimal supersymmetric standard model, MSSM,
the Higgs sector is just two doublets
\cite{Guni,Gun2,DHS,Espa,CPR}. As in
the standard model the requirement on
the electroweak phase transition is that
it be first order so as to avoid washout
of the baryon asymmetry immediately
after the transition; it appears difficult to
avoid this baryon washout in the MSSM \cite{Espa}.

Here we shall discuss the next-to-minimal model,
NMSSM, which has additionally
one singlet Higgs scalar \cite{Guni,Gun2}.
Our treatment is perturbative; we
do not treat the non-perturbative ideas which
it has been suggested \cite{Farr} might
rehabilitate baryogenesis in the
minimal, one Higgs doublet, Standard Model;
and might when fully established
provide a more secure basis for judgement on the various models.
Additionally to the case of electroweak baryogenesis, it is
important to examine the nature of the
electroweak phase transition because
of the many theories which do not create
a net B-L in a phase change at higher
energy; such theories are also menaced by baryon washout.

In the absence of hard information we have to
adopt a hypothesis on the SUSY
breaking scale and on the spectrum of
the particles. We follow a number of
papers of recent years in taking the SUSY
breaking scale, $M_S$, to be of the order of 1 TeV; we take
perfect supersymmetry above that scale \cite{Barb,Frog}.
Then at $M_S$ the quartic scalar couplings are fixed by
the gauge couplings and  two more parameters.
We then use the renormalization group (RG) equations
to run down the quartic couplings to the electroweak scale,
where we investigate the nature of the phase change.
There are also cubic and quadratic supersymmetry
breaking couplings, and there
results a space of variable parameters in which
we investigate what proportion
leads to a strongly first order electroweak phase change,
and so is compatible with electroweak baryogenesis.

There has been quite considerable previous
work on the electroweak phase change
in the MSSM. We are not aware of so much on the
NMSSM . The work of Pietroni \cite{Piet} has
pointed up that the NMSSM, in
contrast to the MSSM, has cubic terms in the
scalar field potential at tree
level leading to the possibility of a potential
barrier in radial directions
even at tree level; and that this diminishes
the importance of the $m^3T$ term,
in the high-T expansion of the T-dependent part,
which has a vital role in
most other theories of the phase change.
That work uses a unitary gauge which
we consider to be not so secure a basis
for the consideration of phase
changes as the Landau gauge which we use \cite{Davl,Doln}.

A further difference is that we include the
$\mu$-term in the superpotential
of the NMSSM; $\mu=0$ is incorporated
as a special case. Though there is a
known naturalness problem with the magnitude
of $\mu$, the often preferred
solution of setting $\mu$ equal to zero raises cosmic
domain wall problems \cite{Absw}. When $\mu\not= 0$
one often used \cite{Davl,Espa,Piet} definition of
the critical temperature, being that temperature
at which the curvature of the
potential at the origin vanishes, can no longer be justified.

We adopt the alternative definition,
being that temperature when tunneling
first becomes possible from the high-T
vacuum to the low-T vacuum having
non-zero Higgs doublets expectation values.
This criterion being well-known
and discussed \cite{Kuzm,Kirz}, has
on occasion actually been used in
calculations \cite{Chvo}, though it is more
difficult to implement than the
curvature criterion. This is an important
change from most previous
calculations, such as those of
Ref.\ \cite{Davl,Espa} in the MSSM and
Ref.\ \cite{Piet} in the NMSSM.
Also this paper has the new feature that it
compares the results from the two criteria
when both can be implemented, that
is in the $\mu=0$ case.

\section{Formalism of the NMSSM}
\label{Nmssm}
The magnitudes of the parameters of the
superpotential are significant for
the electroweak phase transition.
With the usual notation for the quark and lepton
superfields, and summation over generations
understood, the superpotential is
\begin{eqnarray}
\label{eq.sfw}
 W & = & g_{u}Qu^{c}\H{2}+g_{d}Qd^{c}\H{1}+g_{e}Le^{c}\H{1}
\nonumber\\
 & & +\mu\H{1}\H{2}+\lambda\H{1}\H{2}N-\frac{k}{3}N^{3} - rN
\end{eqnarray}
where $\H{1}, \H{2}$ are the Higgs doublet,
and $N$ the singlet, superfield
\footnote{A possible term $N^2$ in the superpotential
can be removed by a field redefinition \cite{Gun2}.
The linear term, $rN$, contributes
terms to the effective potential of the same type as we include
with arbitrary coefficients in the soft supersymmetry-breaking
part of the potential $V_0$ given in Eq.\ (\ref{eq.vs0}).
}.
The corresponding tree level scalar field potential,
below the \s breaking scale, is  \cite{Guni,Gun2}
\begin{eqnarray}
\label{eq.vs0} V_{0} & = & \frac{1}{2}(\lm{1}(\Hd{1}\H{1})^2 +
\lm{2}(\Hd{2}\H{2})^2) + \nonumber\\
 & & (\lm{3}+\lm{4})(\Hd{1}\H{1})(\Hd{2}\H{2}) -
\lm{4}\left| \Hd{1}\H{2}\right|^{2}+ \nonumber\\
 & & (\lm{5}\Hd{1}\H{1} +\lm{6}\Hd{2}\H{2})N^{\star}N+
(\lm{7}\H{1}\H{2}N^{\star2}+h.c.) + \nonumber\\
 & & \lm{8}(N^{\star}N)^2+
(\left|\mu\right|^2+(\lambda\mu^{\star}N+h.c.))
(\Hd{1}\H{1}+\Hd{2}\H{2}) +
\nonumber\\
 & & \m{1}\Hd{1}\H{1}+\m{2}\Hd{2}\H{2} +
\m{3}N^{\star}N-(m_4\H{1}\H{2}N+h.c.)
 -\nonumber\\
 & & \frac{1}{3}(m_5N^3+h.c.)+\frac{1}{2}
(\m{6}\H{1}\H{2}+h.c.)+(\m{7}N^2+h.c.)
\end{eqnarray}
where $$\H{1}^{T} = (\H{1}^{0},\H{1}^{-}), \H{2}^{T}
= (\H{2}^{+},
\H{2}^{0}), \H{1}\H{2} = \H{1}^{0}\H{2}^{0}-
\H{1}^{-}\H{2}^{+}$$
$\H{1}, \H{2}$ and $N$ now denoting pure scalar fields.
The $m$-coefficient
terms comprise all possible soft supersymmetry breaking
terms \cite{Guni,Gun2}.
$V_{0}$ is a function of 10 real scalar fields,
the Higgs doublets and the
singlet being given in terms of these by
\begin{equation} \label{eq.doublets}
\H{1} = \frac{1}{\sqrt{2}}\left( \begin{array}{c}
\p{1} + i\p{2}\\ \p{3} +
i\p{4} \end{array} \right) , \; \H{2} =
\frac{1}{\sqrt{2}}\left(
\begin{array}{c} \p{5} + i\p{6}\\ \p{7} +
i\p{8} \end{array} \right),
N=\frac{1}{\sqrt{2}}(\p{9}+i\p{10})
\end{equation}

For simplicity, and to automatically ensure real VEVs,
we shall follow the usual
practice and take the parameters real.
The boundary values at $M_S$ of the
 quartic couplings are given by \cite{Ekw1}
$$\lm{1} = \lm{2} = \frac{1}{4}(g_2^2+g_1^2),
 \lm{3} = \frac{1}{4}(g_2^2-g_1^2), \lm{4} =
\lambda^2-\frac{1}{2}g_2^2 ,$$

$$\lm{5}=\lm{6}=\lambda^2,\lm{7}=-{\lambda}k,\lm{8}=k^2 $$
and are developed
down to $M_{\rm{Weak}}$ by using the appropriate RG equations
\cite{Ekw1}. The $\lambda$ and k are in principle
free parameters at $M_S$
\cite{Guni,Gun2}. However they are linked to the
one important Yukawa coupling
 \footnote{we are not considering large $\tan\beta$ here}
, $g_t$, by 3 simultaneous RG equations \cite{Ekw2}
in developing from high
energy down to $M_S$; their values there should not
be such that they
correspond to divergent or unnaturally large
values at high energy. From
 Eqs.\ (\ref{eq.vs0}) and (\ref{eq.doublets})
we obtain the zero
temperature potential $V_{0}(\phi)$ as a
function of \mbox{$\phi =
\p{1},\p{2},\ldots,\p{10}$} with the
parameters \lm{i} assumed renormalised
at the electroweak scale. The
vacuum expectation values of the scalars
are to be found at the minimum of
$V_{0}$ and thus satisfy
\begin{equation} \label{eq.mincon} \partial_{i}V_{0}
\left|_{\phi = \langle\phi\rangle} \right. = 0,
\mbox{\hspace{.5in}}
\partial_{i} \equiv
\frac{\partial}{\partial \p{i}}, i = 1,2,\ldots,10
\end{equation}
and also the requirement that $\langle\phi\rangle$
be a minimum.

It is a constraint on our parameters that the
minimum of $V_0$ at zero
temperature is for the VEVs of the Higgs fields
having neutral components only
\begin{equation} \label{eq.vevs} \langle\H{1}\rangle
= \left( \begin{array}{c} v_{1}\\ 0 \end{array}
\right), \; \langle\H{2}\rangle = \left( \begin{array}{c} 0\\
v_{2} \end{array} \right),\langle N\rangle=x
\end{equation}
where $v_{1}$, $v_{2}$ and $x$ are real and
\mbox{$v = \sqrt{v_{1}^{2} + v_{2}^{2}} = 174$ GeV}.
The scalar mass-squared matrix is given by
\begin{equation} \label{eq.massmat} M_{ij}^{2} =
\partial_{i}\partial_{j}V_{0}\left|_{\phi =
\langle\phi\rangle}\right.
\end{equation} and
gives rise to 7 massive physical particles and 3
zero mass would-be Goldstone bosons. The
$\m{1},\m{2},\m{3}$ are standard mass parameters
and are to be specified  in
 terms of the VEVs and the other parameters by
Eq.\ (\ref{eq.mincon}) when
$\phi$ is given by Eq.\ (\ref{eq.vevs}).

We can now discuss the other parameters in $V_0$.

Firstly there are the terms involving $\mu$
which arise from the $\mu$ term in
 the superpotential. This raises the
mu-problem(first noted in the MSSM) in the
context of the NMSSM \cite{Absw,Elli}.
Unlike the $\lambda$ and k parameters of
 the superpotential just discussed there
is no control on $\mu$, which  would
 naturally be expected to take on a
value of the order of magnitude of the
 fundamental scale of the theory, whereas
phenomenologically it should be of
the order of the other electroweak terms.
We do not take the point of view that the
NMSSM can solve this by its having largely
phenomenologically equivalent terms
in the N field and simply setting
$\mu=0$ \cite{Ekw2,Elli}. This can
have its own difficulties when a resulting
$Z_3$ symmetry gives rise to domain
walls \cite{Absw}. We tolerate the mu-problem.
It should be noted that we have
extra $Z_3$ breaking by the phenomenological
term $\m{6}\H{1}\H{2}$. Secondly
 there are the remaining soft parameters
$m_4,m_5,\m{7}$; their provenance
as completing the most general NMSSM
breaking $V_0$ was given in
 Refs.\ \cite{Guni,Gun2}.

Going now to the T-dependent terms in
the effective potential we have the
development, by the usual methods
\cite{Doln} for a temperature sufficiently
 high compared to the masses, given by
\begin{equation}
\label{eq.veff} V(\phi,T) = V_{0}(\phi) -
\frac{N_{eff}\pi^{2}T^{4}}{90} +
V_{2}(\phi,T) + V_{3}(\phi,T) +
V_{ln}(\phi,T)\end{equation}
\begin{equation} \label{eq.v3} V_{3}(\phi,T) =
-\frac{T}{12\pi}
\left[ \sum_{i = 1}^{10}\left[M_{i}^2(\phi,T)
\right]^{\frac{3}{2}} +
4\left[\frac{g_2^{2}}{4}\sum_{i = 1}^{8}
\p{i}^{2}\right]^{\frac{3}{2}} +
2\left[\frac{g_2^{2} + g_1^{2}}{4}\sum_{i = 1}^{8}
\p{i}^{2}\right]
^{\frac{3}{2}}\right]
\end{equation}
\begin{equation} \label{eq.vln} V_{ln}(\phi,T) =
\pm\sum_{i}n_i\frac{M_i^4}{64{\pi}^2}ln
\left[\frac{M_i^2}{c_iT^2}\right]
\end{equation}
In Eq.\ (\ref{eq.vln}) $\pm$ refers to
fermions and bosons respectively, and
 $n_i$ is the number of degrees of
freedom of the particle \cite{Andn}.
 $$V_2=\frac{T^2}{24}V_{trace}$$
where $V_{trace}$ is the trace of the mass squared matrix
\newpage
\begin{eqnarray}
\label{eq.vtrace}
V_{trace} & = & 4\m{1}+4\m{2}+2\m{3}+8\mu^2 +
\nonumber\\
 & & (3\lm{1}+2\lm{3}+\lm{4}+\lm{5}+\frac{3}{4}
(3g_2^2+g_1^2))(2\Hd{1}\H{1})
+ \nonumber\\
 & & (3\lm{2}+2\lm{3}+\lm{4}+\lm{6}+
\frac{3}{4}(3g_2^2+g_1^2)+3g_t^2)
(2\Hd{2}\H{2})+ \nonumber\\
 & & (2\lm{5}+2\lm{6}+4\lm{8})
(2N^{\star}N)+8\lambda\mu(N+h.c.)
+ \nonumber\\
 & &  4\left|(\mu+{\lambda}N)\right|^2 +2\lambda^2
(\left|\H{1}^0\right|^2+\left|\H{2}^0\right|^2)
+4k^2N^{\star}N.
\end{eqnarray}
The last line is the 1-loop contribution of the Higgsinos in
the case where the masses of the winos and binos
are taken large, of order
$M_S$, and is the sum of their squared masses.
The size of the parameters gives
these masses to be significantly less than $M_S$
and thus they should be taken
into account in the 1-loop corrections.
In Eq.\ (\ref{eq.v3}), the first term is
the contribution from the
scalar masses. Here we have not just used
the tree level masses squared, which
are eigenvalues of $\partial_{i}\partial_{j}V_{0}$ as used in
Eq.\ (\ref{eq.vtrace}), but we use the eigenvalues of
$\partial_{i}\partial_{j}(V_{0}(\phi) +
V_{2}(\phi,T))$. It has been shown \cite{Dine,Carr}
that this is equivalent to
correcting the one-loop potential by adding to the loop all ring
diagrams (`daisy diagrams') which makes what would
otherwise be imaginary values of $\left[
M_{i}^{2} \right]^{\frac{3}{2}}$, for some values
of $\phi$, into real values
\cite{Doln,Kirz}. Also in Eq.\ (\ref{eq.v3}), a
correction to the gauge boson
contribution has been included; in the
contribution of the longitudinal
polarisation excitations, there is
suppression due to a temperature dependent
`Debye mass' factor. A simple treatment of
this due to Dine et al.\ \cite{Dine}
is just to drop the longitudinal contribution,
and we follow this prescription.

A technical complication that arises is
that if $\frac{M_i}{T}\ge2.2$ for bosons
or $\frac{M_i}{T}\ge1.8$ for fermions
 then the expansion of Eq.\ (\ref{eq.veff}) breaks down;
 this is important at
 $T=T_{crit}$. We solve this by the method of
 Anderson and Hall \cite{Andn} and
replace the last three terms of Eq.\ (\ref{eq.veff}) by
\begin{equation}\label{eq.delv}
{\Delta}V(\phi)=-\sum_{i}\frac{n_iT^2M_i^2}
{(2\pi)^{\frac{3}{2}}}
\sqrt{\frac{T}{M_i}}e^{-\frac{M_i}{T}}
\left[1+\frac{15T}{8M_i}+\cdots\right]
\end{equation}
in the appropriate high $\frac{M_i}{T}$ region.
Here $n_i$ is the number of
 degrees of freedom of particle i.

\section{Phase Change Calculations}
Calculations with the kinetic equations for the dilution of
baryonic charge just after the phase transition
give a baryon preservation
condition \cite{Dine,Cohn}
\begin{equation} \label{eq.voverT}
\frac{v(T_{crit})}{T_{crit}} \geq \xi
\end{equation}
where $\xi$ varies between about 0.9 and 1.5
according to the gauge and
other couplings of the theory.
We take $\xi = 1$. In practice in various theories
just two criteria have
been used to find the critical temperature,
$T_{crit}$, and authors have
 made a choice of either one or the other;
we use both and compare. One is the
 temperature, $T_0$, at which, for decreasing T,
the curvature of the
effective potential $V(\phi,T)$ at $\phi=0$
(assumed to be the previous global
minimum) first vanishes in the Higgs doublet
neutral field directions. The
other is the temperature, $T_C$, at which the value of V
at a different minimum with a non-zero $v'$ first becomes
the global minimum of $V( v_{1}',v_{2}',
x',T)$; here the prime denotes general values of the
field quantities as
opposed to the specific values  $v_{1},v_{2},x$
proper for the T = 0 physical
 theory. As discussed below only for $\mu=0$ is the
first criterion, $T_{crit}=T_0$, applicable.
 Now the shape of $V( v_{1}',v_{2}',x',T)$
depends on the theory parameters.
 It is the space of variable parameters
which we search to assess in what part,
 and what proportion, of it the criterion
Eq.\ (\ref{eq.voverT}) is satisfied.
These variables are:
\begin{enumerate}
\item $[\mu, \lambda, k]$, occuring in the
superpotential Eq.\ (\ref{eq.sfw});

\item$[\tan\beta \equiv v_{2}/v_{1}, x]$
(together with the known fixed
 value of $v$) specify the VEVs and replace
\m{1}, \m{2}, \m{3};

\item $[M_{ch}]$, the mass of the charged Higgs,
is similarly given by an
 analytic formula, and we take it to
replace $m_4$ as a variable;

\item $[m_5, \m{6}, \m{7}]$; \m{6}, when non-zero,
breaks the $Z_3$ symmetry even in the absence of $\mu$.
\end{enumerate}

It is also necessary to consider the permissible
range for the top Yukawa
coupling, $g_t$. We do not investigate
the case where $\tan\beta$ is very
 large, this restriction implying that
$g_t^2 \gg g_b^2$; there then are
3 simultaneous RG equations for $\lambda,k,g_t$
\cite{Ekw2}. These run from
high energy down to $M_S$, where the
$\lambda,k,g_t$ should not be such that
they correspond to divergent or unnaturally
large values at high energy. The
first limitation that this induces is
$g_t(M_S)\le 1.06$ which corresponds
to $g_t(m_t)< 1.12$. The experimental
value of $m_t(m_t)$, being greater than
 about 150 GeV, induces a lower limit on
$g_t$ through the relation
$m_t = g_{t}v\sin\beta$, where $v$ = 174 GeV,
resulting in $g_t(m_t)>0.85$.

Within the above small available range for $g_t$
we select $g_t(M_S) = 1$ and
we now discuss the relevant ranges of
the above variables (i)-(iv):
\begin{enumerate}
\item $[\mu, \lambda, k]$: We start by
noting that the presence of $\mu$,
important fundamentally,
has an interesting consequence for the
minimum of V at non-zero T. We discuss
this in the high T expansion. As displayed above,
Eq.\ (\ref{eq.vtrace}), $V_2$
is a function of the
fields with the only linear terms being
those in $(\lambda\mu N+ h.c.)$.
 This means that for any non-zero T
the origin $\phi_i = 0$ cannot be a
 turning value and so cannot be a minimum for $\mu \neq 0$.
We are in a situation different from that in the MSSM
but similar to that discussed by Choi and Volkas \cite{Chvo}
 in a different theory: the false vacuum is not
at the origin, and the
transition to the true vacuum is more
complicated but amenable to
investigation. It follows that only for
the special case $\mu = 0$ is the
critical temperature criterion
of vanishing curvature at the origin,
$T_{crit}=T_{0}$, applicable. We have investigated the cases
 $\lambda\mu=0$, $\pm20$ GeV, $\pm40$ GeV, $\pm60$ GeV.
The RG equations in $\lambda,k,g_t$ also limit
the ranges of $\lambda$ and
$k$, which we find roughly require
$\sqrt{\lambda^{2}+k^{2}}/g_t$ to be
less than about 0.8
\cite{Ekw2}.
With our value $g_t(M_S) = 1$
we have investigated the
three cases $(\lambda,k)=$ (0.65, 0.1); (0.1, 0.65); (0.5, 0.5).

\item $[\tan\beta, x]$: The value of $\tan\beta$ is associated
with that of the Yukawa
coupling and the top mass through $m_t = g_tv\sin\beta$.
We have investigated
values of $\tan\beta$ in the range 1.6 to 3.0 in steps of 0.2
which, with our value of $g_t$,
corresponds to a running top mass of 150 to 173 GeV.
The parameter $x$ is
rather free; the values considered are:
$x = 0.1v$, $0.5v$, $v$, $2v$.

\item $[M_{ch}]$: Accepting the interpretation
of the CLEO $b \to s\gamma$ rate as requiring
that $M_{ch}>200$ GeV, we have searched in the region
$200<M_{ch}<400$ GeV.

\item $[\m{6},m_5,\m{7}]$ The
range of searches of these variables has been
$$-6 < \frac{\m{6}}{(50 \; GeV)^2} < 6,\quad -6 <
\frac{\m{7}}{(50 \; GeV)^2} < 6,
\quad -1 < \frac{m_5}{(60 \; GeV)} < 1 .$$
\end {enumerate}
In the ranges indicated above for $M_{ch},m_5,\m{6}$ and $\m{7}$,
we have conducted
a coarse-grained search on a multidimensional grid over the
parameter space.

The first computing task is to restrict
the parameter space to satisfy experimental constraints at
T = 0. The conditions imposed are:
\begin{enumerate}
\renewcommand{\labelenumi}{(\roman{enumi})}
\item the masses are real,

\item the minimum of V at the specified  $\tan\beta,v$ (= 174 GeV)
and $x$ is a global minimum in the space  $v_{1}',v_{2}',x'$,

\item the charginos and neutralinos have masses greater
than 45 GeV and 30 GeV respectively, and

\item the mass of the lightest Higgs scalar is greater than 65 GeV.

\end{enumerate}

We call the set of these acceptable parameter sets the
 {\it basis space}. These conditions on the basis space are
more restrictive than those imposed in
our conference report \cite{Moor}.
This results in a much smaller basis space
than in \cite{Moor} and a
much greater proportion of baryon preserving cases.

The experimental limitation, (iv), on
the lightest Higgs mass does not severely
limit the basis space; it cuts out
only about 1/3 of the sets which would
otherwise pass into the basis space,
whereas the first three conditions have a more severe effect.

\section{Results and Conclusions}
We need to find what proportion of the above
basis space satisfies the baryon
number preserving condition Eq.\ (\ref{eq.voverT}).
We shall take this as a
measure of the ease of baryon preservation in the NMSSM theory.

We have examined an initial grid of 105,840
parameter sets. Of these
 1,760, being 1.66\%, pass through the
tests (i)-(iv) of Section 3 for
producing an appropriate T = 0 mass spectrum and
form the basis space. We have
then tested each parameter set in that space for
whether it passes the
baryon preservation test, with the critical
temperature defined as being that
 where tunneling from the higher temperature
vacuum to the low temperature
vacuum first becomes possible, as described in Section 3.

The breakdown of these numbers into sectors
of different $\mu, \lambda, k$
and $x$, is given in Table 1. It is clear,
 both from the overall result and from the
breakdown, that the condition of
baryon number preservation just
after the electroweak phase change does not
impose a significant extra constraint on
the parameters of the NMSSM from the present,
statistical, point of view.
Overall 50\% of the basis space passes
the test of baryon preservation.
Despite the large number of parameter
sets in the initial grid,
the sample of ultimately successful
cases is around 900,
too small to permit a meaningful
delineation of the acceptable
regions of our 9-dimensional
parameter space. In the particular
case of the parameters $\mu$,
$\lambda$, $k$ of the superpotential,
Table 1 gives an indication of the
variation of the successful
proportion with $\lambda$ and $k$;
for $\mu$ the criteria select
a basis space with a very strong
bias towards negative $\mu$
(relative to $x$) and, after that
selection, the baryon preservation
condition gives a rather constant
proportion over different values of
$\mu$ and so does not select further.
The 65 GeV lower bound we have placed on the mass
of the lightest Higgs scalar
is ultra-conservative, since it comes from analyses of
experiments performed within the SM and the MSSM.
In the NMSSM, with
its extra mixing and different couplings,
lower mass Higgs scalars
could have escaped detection \cite{Kw}.
All parameter sets in our
basis space yield a lightest Higgs
scalar with a mass less than
120 GeV. The tendency
is for lower mass lightest Higgs scalars
to be associated with a greater
proportion of baryon preserving cases
(80\% for a mass of 70 GeV but only 20\% for 120 GeV).

In Table 2 we show some results for the
special case $\mu=0$. Of chief
 interest is the comparison between the zero curvature at
 $\langle \phi \rangle =0$ criterion and the
tunneling criterion for the
 critical temperature. Though there are
some case by case differences the two
 criteria are statistically nearly the same;
to our knowledge this is the
 first time such a detailed comparison has been made.
It was found that the depth of the broken
symmetry minimum relative to
the value of V in the neighbourhood of
the origin changed rapidly
with temperature. As the perturbative
effective potential formalism
suffers from well-known problems near
the origin particularly
\cite{Doln,esp,kaj}, the tunneling
criterion is preferable to the frequently used
curvature criterion, being less dependent on
the value precisely at the origin.

Our conclusions for the case $\mu = 0$
broadly agree with previous work on
this case \cite{Piet}, although our
formalism differs, e.~g.\
our gauge choice produces different
T-dependent contributions to the
scalar mass matrix. To treat the hitherto
uninvestigated case of
$\mu \neq 0$, we have had to change
from the vanishing curvature
criterion to the tunneling criterion
for the critical temperature;
we have found that values of $\mu$
up to about 100 GeV are acceptable,
with well-chosen values of the other
superpotential parameters
($\lambda, k$). Unlike the Standard Model
or the MSSM, the NMSSM can
easily coexist with baryon number
preservation. The essential
difference is that cubic terms can
already be present in the
T = 0 potential, so that the T-dependent
terms modify but are
not solely reponsible for the existence
of a minimum away from
the origin, $v = 0$, at the critical temperature.

\section*{Acknowledgements}

We should like to thank P. L. White
and D. G. Sutherland for very
useful and stimulating discussions.

\newpage
\begin{table}
\caption{Showing the proportion of the
basis space of parameter sets
which satisfy the B( baryon)-preservation
criterion, as a function of
 $\lambda,k$ and of the T = 0 value of
$x$ ($\equiv\langle N \rangle $).
 In the basis space the T = 0 criteria
are satisfied. The values of $\mu$
covered are given by $60$ GeV $>\lambda\mu> -60$ GeV.}
\begin{displaymath}
\begin{array}{|cc|c|c|}
\hline
\lambda & k & x/v &
\begin{array}{c}\rm{B\;preserving}\\
\%\rm{-age}\;\rm{of\;basis}\end{array}\\ \hline
0.65 & 0.1 & \begin{array}{c}0.1\\0.5\\1.0\\2.0\end{array} &
\begin{array}{c}40\\66\\82\\59\end{array}\\ \hline
0.1 & 0.65 & 1.0 & 50 \\ \hline
0.5 & 0.5 & 1.0 & 26 \\ \hline
\end{array}
\end{displaymath}
\label{SU5Quarks}
\end{table}

\begin{table}
\caption{$\mu$=0: Comparing the results from
the criterion of vanishing
curvature of V at $ \langle \phi \rangle = 0,
\rm{T_{crit}=T_{0}}, $
with the tunneling criterion, $\rm{T_{crit}=T_{C}}$.
The baryon preserving
 proportion of the basis space and
the average value of the critical
temperature, $ \rm{T_{crit}} $, are shown for both cases .}
\begin{displaymath}
\begin{array}{|ccc|c|c|c|c|}
\hline
\lambda & k & x/v &
\begin{array}{c}\rm{B\;preserving}\\ \%\rm{-age}
\; \rm{for}\\ T_{crit}=T_{0}\end{array} &
\begin{array}{c}\rm{Average\;of}\\\rm{T_{crit}=T_{0}}
\end{array} &
\begin{array}{c}\rm{B\;preserving}\\ \%\rm{-age}
\; \rm{for}\\ T_{crit}=T_{C}\end{array} &
\begin{array}{c}\rm{Average\;of}\\ \rm{T_{crit}=T_{C}}
\end{array}\\ \hline
0.65 & 0.1 & 1.0 & 80 & 70\; \rm{GeV} & 79 & 71\;
\rm{GeV} \\ \hline
0.5 & 0.5 & 1.0 & 16 & 82\;
\rm{GeV} & 11 & 80\; \rm{GeV} \\ \hline
\end{array}
\end{displaymath}
\label{mueq0}
\end{table}

\end{document}